\documentclass[12pt]{iopart}
\usepackage{graphicx}
\usepackage{dcolumn}
\usepackage{bm}

\newcommand{\ba}{\begin{eqnarray}}
\newcommand{\ea}{\end{eqnarray}}
\newcommand{\ban}{\begin{eqnarray*}}
\newcommand{\ean}{\end{eqnarray*}}
\newcommand{\be}{\begin{equation}}
\newcommand{\ee}{\end{equation}}
\newcommand{\bd}{\begin{displaymath}}
\newcommand{\ed}{\end{displaymath}}

\newcommand{\n}[1]{\label{#1}}
\newcommand{\nn}[1]{(\ref{#1})}
\newcommand{\non}{\nonumber}
\newcommand{\Eq}[1]{(\ref{#1})}

\newcommand{\x}{{\bi x}}

\newcommand{\pa}{\partial}

\newcommand{\lap}{\bigtriangleup}

\newcommand{\hh}{\, ,\hspace{0.5cm}}
\newcommand{\hhh}{\, ,\hspace{0.2cm}}

\begin{document}


\title{Gravitational field of charged gyratons}

\author{Valeri P. Frolov$^{1}$ and Andrei Zelnikov$^{1,2}$}

\address{$^{1}$Theoretical Physics Institute, University of Alberta,
Edmonton, Alberta, Canada, T6G 2J1}
\address{$^{2}$Lebedev Physics Institute,
  Leninsky prospect 53, 119991, Moscow Russia}
\ead{frolov@phys.ualberta.ca, zelnikov@phys.ualberta.ca}
\begin{abstract}
We study relativistic gyratons which carry an electric charge. The 
Einstein-Maxwell equations in arbitrary dimensions are solved exactly in the case of a
charged gyraton propagating in an asymptotically flat metric. 
\end{abstract}

\maketitle

\section{Introduction}

Possible creation of mini black holes in the scattering of highly
ultra-relativistic particles attracted recently a lot of attention.
For such a process the energy of particles in their center of mass must
be of order of or higher than the fundamental (Planckian) mass $M_*$.
In the "standard" quantum gravity this energy is very high, $10^{28}$~eV.   
This makes it difficult to expect that this, theoretically interesting
and important, process will be observed in a near future. The
fundamental mass scale might be much smaller if recently proposed models
with large extra dimensions are valid. The fundamental energy
which is often discussed in these models is of order of a few TeV.
This opens an appealing possibility of mini black hole
creation in the near future collider and cosmic ray experiments.

In order to estimate the cross-section of the mini black hole production one
needs to know the gravitational field generated by an ultra-relativistic
particle. To calculate this field it is sufficient to boost the Schwarzschild
metric keeping the energy fixed and take the limit when the boost parameter
$\gamma=1/\sqrt{1-v^2/c^2}$ becomes very large (the so called Penrose limit).
The corresponding metric in the 4-dimensional spacetime was obtained by
Aichelburg-Sexl \cite{AiSe}. It can be easily generalized to the case of a
spacetime with arbitrary number of dimensions $D=n+2$ where it has the form
\be\n{as1}
ds^2=-2\,du\,dv+d{\bi x}^2+\Phi(u,{\bi x}) du^2\, ,
\ee 
\be\n{as2}
\Phi(u,{\bi x})= \kappa \sqrt{2} E \delta(u) f(r)\, ,
\ee
\be\n{as3}
f(r)= \displaystyle {g_n\over r^{n-2}}\hh 
g_n={\Gamma( { n-2\over 2})/ (4\pi^{n/2})}
\, .
\ee
Here $d\x^2$ is the metric in the $n$-dimensional Euclidean space $R^n$,
$r^2={\bi x}^2$, and $\kappa=16\pi G$. In the 4-dimensional spacetime 
$f(r)=-(1/2\pi)\ln r$.

The constant $E$ which enters  (\ref{as2}) is the energy. One can easily smear
the $\delta$-like profile of the energy distribution in (\ref{as2}) by
substituting a smooth function $\varepsilon(u)$ instead of $E\delta(u)$. In
what follows we assume that the function $\varepsilon(u)$ (and similar
functions which appear later) vanishes outside a finite interval $u\in
[u_1,u_2]$. For this smooth distribution, $\varepsilon(u)$ has the meaning of
the energy density, so that the total energy of such an object moving with the
velocity of light is 
\be\non
E=\int_{u_1}^{u_2} du\,\varepsilon(u) \, .
\ee
For example such a gravitational field is produced by a pulse of light which has
the finite energy $E$ and the finite duration (in time $u$) $L=u_2-u_1$.

Studies of the gravitational field of beams and pulses of light have a long
history. Tolman \cite{To} found a solution in the linearized approximation.
Peres \cite{Pe1,Pe2} and Bonnor \cite{Bo} obtained exact solutions of the
Einstein equations for a pencil of light. These solutions belong to the class
of  $pp-$waves. The $pp-$wave solutions in four dimensions are described
in detail in the book \cite{Steph}. The four-dimensional $pp-$waves created by
an ultrarelativistic charge or a spinning null fluid  have been found by Bonnor
\cite{Bo2,Bo3}. The higher-dimensional generalization of these solutions to the
case where the beam of radiation carries angular momentum has been found
recently  \cite{FrFu:05,FrIsZe:05}. Such solutions correspond to a pulsed beam
of radiation with negligible radius of cross-section, finite duration $L$ in
time, and which has finite both energy $E$, and angular momentum $J$. An
ultra-relativistic source with these properties was called {\em a gyraton}. 

In the present paper we continue studying properties of gyratons. Namely we
obtain a solution for the electrically charged gyraton and describe its
properties.


\section{Ansatz for metric and electromagnetic field}

Our starting point is the following ansatz for the metric in the $D=n+2$
dimensional spacetime
\be\n{met}
ds^2=d\bar{s}^2+2\left(a_u du+a_a dx^a\right)\,du\, .
\ee
Here
\be\n{flat}
d\bar{s}^2=-2\,du\,dv+ d{\bi x}^2
\ee 
is a flat $D-$dimensional metric, and $a_u=a_u(u,x^a)$, $a_a=a_a(u,x^a)$. The
spatial part of the metric $d{\bi x}^2=\delta_{ab}dx^adx^b$ in the
$n-$dimensional space $R^n$ is flat. Here and later the Greek letters $\mu$,
$\nu \ldots$ take values $1, \ldots, D$, while the Roman low-case letters $a$,
$b, \ldots$ take values $3, \ldots, D$. We denote
\be\non
a_\mu=a_u\delta_\mu^u+a_a\delta_\mu^a.
\ee
The form of the metric \nn{met} is invariant under the following coordinate
transformation
\be\n{gauge}
v\to v+\lambda(u,{\bi x})\hhh
a_{\mu}\to a_{\mu}-\lambda_{,\mu}\, .
\ee
This transformation for $a_{\mu}$ reminds the gauge transformation for the
electromagnetic potential. The quantity 
\ban
f_{\mu\nu}=\partial_{\mu}a_{\nu}-\partial_{\nu}a_{\mu},
\ean
which is the gravitational analogue of the electromagnetic strength tensor, is
gauge invariant.

The metric \nn{met} admits a parallelly propagating null Killing
vector  $l=l^{\mu}\pa_{\mu}=\pa_v$. It is the most general
D-dimensional null Brinkmann metrics \cite{Brinkmann:25} with flat
transverse space. Sometimes they are called $pp$-wave metrics
\cite{Tseytlin:95}. The metric for the gravitational field of a
relativistic gyraton with finite energy and internal angular momentum
(spin) is of the form \nn{met} \cite{FrFu:05,FrIsZe:05}.

Let us denote
\be\non
l_{\mu}=-\delta_\mu^u.
\ee
Then the flat metric $d\bar{s}^2$ and the metric \nn{met} are related as
\ban
\bar{g}_{\mu\nu}=g_{\mu\nu}-l_{\mu}a_{\nu}-l_{\nu}a_{\mu}\, ,
\ean
\ban
\bar{g}^{\mu\nu}=g^{\mu\nu}+l^{\mu}a^{\nu}+l^{\nu}a^{\mu}
+l^{\mu}l^{\nu}a^{\epsilon}a_{\epsilon}\, .
\ean
In the last relation, $\bar{g}^{\mu\nu}$ is the metric inverse  to
$\bar{g}_{\mu\nu}$. The operations with the indices of the quantities which
enter its right hand side are performed by using the metric $g_{\mu\nu}$ and
its inverse $g^{\mu\nu}$. It is easy to check that
\be\n{al1}
l^{\epsilon}a_{\epsilon}=l^{\epsilon}f_{\epsilon\mu}=0\, .
\ee
For the metric ansatz \Eq{met} all local scalar invariants constructed from the
Riemann tensor and its derivatives vanish \cite{FrIsZe:05}. This property can
be proved off-shell using \Eq{al1} and the fact that the the Riemann tensor is
aligned with the null Killing vector $l_\alpha$. The tensor is called aligned
to the the vector $l_{\alpha}$  if it can be written as a sum of terms, where
each term contains as a factor at least one vector $l_{\alpha}$. Vanishing of
all scalar invariants constructed from the curvature and its derivatives is
important in the proof of absence of quantum and $\alpha'$ corrections to the
gravitational shock waves  \cite{HorowitzSteif:90,AmatiKlimcik:89}. One can
expect that for the same reason the gyraton metric also does not have local
quantum corrections.

Now let us consider the electromagnetic field in the spacetime \nn{met}. We
choose its vector potential $A_{\mu}$ in the form
\be\n{empot}
A_\mu=A_u\delta_\mu^u+A_a\delta_\mu^a,
\ee
where the functions $A_u$ and $A_a$ are independent of the null
coordinate $v$. 
The electromagnetic gauge transformations
\ban
A_\mu\to A_\mu-\Lambda_{,\mu}
\ean
with $\Lambda=\Lambda(u,{\bi x})$ preserve the form of the potential
\nn{empot}.

The Maxwell strength tensor for this vector potential is
\ban
F_{\mu\nu}=\partial_{\mu}A_{\nu}-\partial_{\nu}A_{\mu}\, .
\ean
Being written in the covariant form, it does not depend on the metric and hence
it is the same both for the metric \nn{met} and for the background flat metric
\nn{flat}, $\bar{F}_{\mu\nu}=F_{\mu\nu}$. The Maxwell tensors with the upper
indices for these spaces are different and are related as follows
\ban
\bar{F}^{\mu\nu}=F^{\mu\nu}+l^{\mu}a_{\sigma}F^{\sigma\nu}
-l^{\nu}a_{\sigma}F^{\sigma\mu} \hh
\bar{F}_{\mu}{}^{\nu}=F_{\mu}{}^{\nu}+l^{\nu}a_{\sigma}F_{\mu}{}^{\sigma}\,
.
\ean
The quantities $A_{\mu}$ and $F_{\mu\nu}$ obey the relations
\ba\n{orthog}
l^{\epsilon}A_{\epsilon}=l^{\epsilon}F_{\epsilon\mu}=0\, ,
\ea
which imply that
\ban
\bar{F}_{\mu}^{\ \ \alpha}\bar{F}_{\alpha\nu}={F}_{\mu}^{\ \
\alpha}{F}_{\alpha\nu}\, .
\ean


\section{Field equations}

The Einstein-Maxwell action in higher dimensions reads
\ban
S={1\over \kappa}\int d^D\x\sqrt{|g|}\left[R-{1\over
4}F^{\mu\nu}F_{\mu\nu}+A_{\mu}J^{\mu}\right].
\ean
Here $\kappa=16\pi G$ and $G$ is the gravitational coupling constant in
$D-$dimensional spacetime. For this form of the action both the metric
$g_{\mu\nu}$ and the  vector potential $A_{\mu}$ are dimensionless 
\cite{Ortin}.

The stress-energy tensor for the electromagnetic field is
\be\n{set}
T_{\mu\nu}={1\over \kappa}\left[ F_{\mu}{}^{\epsilon}F_{\nu\epsilon}-{1\over 4}g_{\mu\nu}
F_{\epsilon\sigma}F^{\epsilon\sigma}\right].
\ee
For the potential \nn{empot} it takes the form
\ba
T_{\mu\nu}={1\over \kappa}\left[ F_{\mu}{}^{\epsilon}F_{\nu\epsilon}-{1\over 4}g_{\mu\nu}
{\bi F}^2 \right]
={1\over \kappa}\left[ \bar{F}_{\mu}{}^{\epsilon}\bar{F}_{\nu\epsilon}-{1\over
4}\left(\bar{g}_{\mu\nu}+l_{\mu}a_{\nu}+l_{\nu}a_{\mu}\right)
{\bi F}^2 \right],
\ea
where
\be\non
{\bi F}^2=F_{\epsilon\sigma}F^{\epsilon\sigma}=
\bar{F}_{\epsilon\sigma}\bar{F}^{\epsilon\sigma}=
F_{ab}F^{ab}\, .
\ee
The trace of the stress-energy tensor
\be\n{tr}
T^{\mu}_{\mu}={4-D\over 4\kappa}{\bi F}^2
\ee
vanishes in the 4-dimensional spacetime. This is a consequence of the
conformal invariance of 4-dimensional Maxwell theory. Since 
$F_{a}{}^{\epsilon}F_{b\epsilon}= F_{a}{}^{c}F_{bc}=
\bar{F}_{a}{}^{c}\bar{F}_{bc}$ and $g_{ab}=\bar{g}_{ab}=\delta_{ab}$
one has the following expression for the partial trace 
\be\n{ptr}
\delta^{ab}T_{ab}={6-D\over 4\kappa} {\bi F}^2\, .
\ee

Our aim now is to find solutions of the system of Einstein-Maxwell
equations
\be\n{ein}
R_{\mu\nu}-{1\over 2}g_{\mu\nu} R={1\over 2}\kappa T_{\mu\nu}\, ,
\ee
\be\n{max}
F_{\mu}{}^{\nu}{}_{;\nu}=J_\mu\, ,
\ee
for the adopted field ansatz \nn{met} and \nn{empot}. The
stress-energy tensor $T_{\mu\nu}$ is given by \nn{set}.

Direct calculations \cite{FrIsZe:05} show that for the metric \nn{met} the
scalar curvature vanishes $R=0$ and the only nonzero components of the Ricci
tensor are
\ban
R_{ua}&=&{1\over 2}f_{ab}{}^{,b} ,\\
R_{uu}&=&-(a_u)^{,a}_{,a}+{1\over 4}f_{ab}f^{ab}+\partial_u
(a_a{}^{,a})\, .
\ean

Since $R=\delta^{ab}R_{ab}=0$ by using relations \nn{tr} and \nn{ptr}
one obtains
\be\non
{\bi F}^2=0\, ,
\ee
and hence $F_{ab}=0$. Thus in a proper gauge the transverse components of the
electromagnetic vector potential vanish, $A_a=0$. Let us denote 
${\cal A}=A_{u}$, then the only non-vanishing components of
$F_{\mu\nu}$ and $T_{\mu\nu}$ are
\be\n{Fua}
F_{ua}=-F_{au}=-{\cal A}_{,a}\hh
T_{uu}={1\over\kappa} (\nabla {\cal A})^2\, .
\ee
where $(\nabla {\cal A})^2=\delta^{ab}{\cal A}_{,a}{\cal A}_{,b}$.

Thus the requirement that the electromagnetic field is consistent with the
Einstein equations for the gyraton metric ansatz \Eq{met} implies that the
vector potential $a_\mu$ can be chosen  $a_\mu\sim l_\mu$, i.e., to be aligned
with the null Killing vector. In this case $F_{\mu\nu}$ and all its covariant
derivatives are also aligned with $l_\mu$. Together with the orthogonality
conditions \Eq{orthog} these properties can be used to prove that all local
scalar invariants constructed from the Riemann tensor, the Maxwell tensor
\Eq{Fua} and their covariant derivatives vanish. This property generalizes the 
analogous property for non-charged gyratons \cite{FrIsZe:05} and gravitational
shock waves. This property can also be used to prove that the charged gyraton
solutions of the Einstein-Maxwell equations are also exact solutions of any
other nonlinear electrodynamics and the Einstein equations 
\footnote{Let us denote ${\cal F}=F_{\mu\nu}F^{\mu\nu}$ and 
${\cal K}=\e^{\mu\nu\alpha\beta}F_{\mu\nu}F_{\alpha\beta}$, then in the
nonlinear electrodynamics the Lagrangian for the field $A_\mu$ is of the form
$L=\cal{F}+\cal{L}(\cal{F,K})$,  where
${\partial\cal{L}\over\partial\cal{F}}\Big|_{{\cal{F}}=0}
={\partial\cal{L}\over\partial\cal{K}}\Big|_{{\cal{K}}=0}=0$. Since for the
gyraton ansatz ${\cal{F}}={\cal{K}}=0$ the non-linear term ${\cal L}$ in the
Lagrangian does not affect the electromagnetic field equations. So that the
solutions for the non-linear electrodynamics coincide with the solutions of the
standard Maxwell equations. We are grateful to David Kubiznak for drawing our
attention to this fact.}.

Eventually the Einstein equations reduce to the following two sets of
equations in $n-$dimensional flat space $R^n$
\be\n{eu}
(a_u)^{,a}_{,a}-\partial_u (a_a{}^{,a})={1\over
4}f_{ab}f^{ab}-{1\over 2}(\nabla {\cal A})^2\, ,
\ee
\be\n{ea}
f_{ab}{}^{,b}=0\hh f_{ab}=a_{b,a}-a_{a,b}\, .
\ee
We are looking for the field outside the region occupied by the gyraton, where
$J_{\mu}=0$. 
The Maxwell equations then reduce to the relation
\be\n{m}
\lap {\cal A}=0\, .
\ee
Here $\lap$ is a flat $n-$dimensional Laplace operator.

If the gyraton carries an electric charge 
\ba\non
Q&=&{1\over\kappa}\int_{\Sigma} J^{\mu}d\Sigma_{\mu},
\ea
then by using the Stoke's theorem it can be written as
\ba\non
Q&=&{1\over\kappa}\int_{\partial\Sigma} F^{\mu\nu}d\sigma_{\mu\nu}.
\ea
Thus the electric charge of the gyraton is determined by the total flux of the
electric field across the surface $\partial\Sigma$ surrounding it. 

Let us choose $\Sigma$ to be $(D-1)-$dimensional region on the surface
$v=\mbox{const}$ with the boundary $\partial\Sigma$ which consists of the
surface of the cylinder $r=$const, $\partial\Sigma_r$,  and two disks of the
radius $r$, $\partial\Sigma_1$ and $\partial\Sigma_2$, located at $u=u_1$ and 
$u=u_2$, respectively. Since $F^{vu}=0$ the fluxes through the boundary disks
vanish. Thus we have
\ba\n{Q1}
Q&=&{r^{n-1}\over\kappa}\int_{u_1}^{u_2} du\int d\Omega_{n-1} F_{ur}(r,{\bi n}).
\ea
Here $\bi n$ is the point on the unit $(n-1)$-dimensional sphere and 
$d\Omega_{n-1}$ is its volume element. 
For the linear distribution of charge one has
\ba\n{Q2}
Q\equiv\int_{u_1}^{u_2} du\,\rho(u).
\ea
where $\rho(u)$ is a linear charge density. 
By comparing \Eq{Q1} and \Eq{Q2} one can conclude that in the asymptotic region 
$r\to\infty$ the following relations are valid
\ban
F_{ur}\approx\left\{\begin{array}{ll}
{\displaystyle\kappa(n-2)g_n\rho(u)\over \displaystyle r^{n-1}},&~\mbox{for}~n>2,\\
{}&\\
{\displaystyle\kappa\rho(u)\over \displaystyle 2\pi r},&~\mbox{for}~n=2.
\end{array}\right.
\ean
Here $g_n$ is defined by \Eq{as3}.

Now we return to the Einstein equations.
The combination which enters the left hand side of \nn{eu} is 
invariant under the transformation \nn{gauge}. One can use this
transformation to put $a_a{}^{,a}=0$. We shall use this "gauge" choice
and denote $a_u={1\over 2}\Phi$ in this "gauge". The equations \Eq{eu}-\Eq{ea}
take the form
\ba\n{Phi}
&&\lap \Phi=-j\hh j=j_f+j_{{\cal A}} \hh j_f\,=\,-{1\over 2}f_{ab}f^{ab} \hh
j_{{\cal A}}=(\nabla {\cal A})^2\, ,\\
&&\n{lapa} \lap a_a=0\, .
\ea

The set of equations  \nn{m}, \nn{Phi}, and \nn{lapa}  determines the
metric
\be\non
ds^2=d\bar{s}^2+\Phi du^2+2 a_a dx^a\,du 
\ee
and the electromagnetic field ${\cal A}$ of a gyraton. (Let us
emphasize again that these equations are valid only outside the
region occupied by the gyraton.)

The equations  \nn{m} and \nn{lapa} are linear equations in an Euclidean
$n$-dimensional space $R^n$. The first one is the equation for the gravitomagnetic
field $a_\mu$ and it formally coincides with the equation for the
magnetic field. The second is an electrostatic equation for the electric 
potential ${\cal A}$. The last (gravitoelectric) 
equation \Eq{Phi} is a linear equation  
in the Euclidean space for the gravitoelectric potential $\Phi$ which
can be solved after one finds solutions for $f_{ab}$ and ${\cal A}$. Thus for a
chosen ansatz for the metric and the electromagnetic fields, the solution of
the Einstein-Maxwell equations in $D-$dimensional spacetime reduce to linear
problems in an Euclidean $n-$dimensional space ($n=D-2$). 

\section{Solutions}

Let us discuss first the scalar (electrostatic) equation (\ref{m}). 
Its general solution for a point-like charge distribution in the
$n-$dimensional space can be written in the form (see e.g. \cite{RuOr}
and references therein)
\be\n{epot}
{\cal A}=\sum_{l=0}^{\infty} \sum_q {{\cal Y}^{lq}\over r^{n+2l-2}}\, .
\ee
Here
\be\non
{\cal Y}^l=C_{c_1\ldots c_{l}}  x^{c_1}\ldots x^{c_{l}}\, ,
\ee
where $C_{c_1\ldots c_{l}}$ is a symmetric traceless rank-$l$
tensor. The index $q$ enumerates linearly
independent components of coefficients $C_{c_1\ldots c_{l}}$. It
takes the value from 1 to $d_0(n,l)$, where
\be\non
d_0(n,l)={(l+n-3)! (2l+n-2)\over l! (n-2)!}
\ee
is the total number of such independent components for given $n$ and
$l$. For the gyraton solution  $C_{c_1\ldots c_{l}}$ are arbitrary
functions of $u$.

In a similar way,  a solution of the  gravitomagnetic equations
\nn{ea} for a point-like source can be written as follows (see
\cite{RuOr} and references therein)
\be\n{an}
a_a=\sum_{l=1}^{\infty} \sum_{q} {{\cal Y}^{lq}_a\over r^{n+2l-2}}\, .
\ee
Here
\be\non
{\cal Y}^l_a=C_{abc_1\ldots c_{l-1}} x^b x^{c_1}\ldots x^{c_{l-1}}\, ,
\ee
where $C_{abc_1\ldots c_{l-1}}$ is a $(l+1)$-th-rank constant tensor
which possesses the following properties: it is antisymmetric under
interchange of $a$ and $b$, and it is traceless under contraction of any
pair of indices \cite{RuOr}.
Again, we use an index $q$ to enumerate different linearly independent
vector spherical harmonics. The total number of these harmonics for
given $l$ is  \cite{RuOr}
\be\non
d_1(n,l)={l(n+l-2)(n+2l-2)(n+l-3)!\over (n-3)! (l+1)!}\, .
\ee 
In the gyraton solution the coefficients $C_{abc_1\ldots
c_{l-1}}$  are arbitrary functions of the retarded time $u$. The
functions $a_a$ obey the following "guage" fixing condition
$a_{a}^{\ ,a}=0$.

In order to obtain a general solution of the gravitostatic equation
\nn{Phi} it is convenient to write $\Phi$ in the form
\be\non
\Phi=\varphi+\psi\, ,
\ee
where $\varphi$ is a general solutions for a point-like source and
$\psi$ is a special solution of the inhomogeneous equation
\be\n{J}
\lap \psi=-j
\ee
in the absence of a point-like source. A general solution $\varphi$
can be written in the form similar to \nn{epot}, while the special
solution $\psi$ can be presented as follows
\be\n{psi}
\psi(u,{\bi x})= \int d{\bi x}' {\cal G}_n({\bi x},{{\bi x}'}) j(u,{\bi x}')\, .
\ee
Here ${\cal G}_n({\bi x},{{\bi x}'})$ is the Green's function for the
$n-$dimensional Laplace operator
\be\n{gf}
\lap {\cal G}_n({\bi x},{{\bi x}'})
=-\delta ({\bi x}-{{\bi x}'})\, ,
\ee
which can be written in the following explicit form
\be\n{gf2}
{\cal G}_2({\bi x},{{\bi x}'})=-{1\over 2\pi}\ln |{\bi x}-{\bi x}'|\, ,
\ee
\be\n{gfn}
{\cal G}_n({\bi x},{{\bi x}'})=
{g_n \over |{\bi x}-{{\bi x}'}|^{n-2}}
\hhh n>2 \, ,
\ee
where $g_n$ is given by \nn{as3}.

\section{Higher-dimensional charged gyratons}

We consider now special solutions for charged gyratons in a spacetime
with the number of dimensions $D>4$. ($4-$dimensional case will be
discussed in the next section.) These solutions are singled out by the
property that in the harmonic decomposition of the functions
$\varphi$, $a_a$, and ${\cal A}$ only the terms with the lowest
multipole momenta are present
\be
\varphi={\varphi_0\over r^{n-2}} \hh
{\cal A}={ {\cal A}_0\over r^{n-2}} \hh
a_a={a_{ab} x^b\over r^n}  \, ,
\ee
where $\varphi_0$, ${\cal A}_0$, and $a_{ab}$ are functions of $u$. It
can be shown \cite{FrIsZe:05} that
\be
\varphi_0(u)=\kappa \sqrt{2} g_n\varepsilon(u)\hhh
a_{ab}(u)=\kappa (n-2) g_n j_{ab}(u)\, ,
\ee
where $\varepsilon(u)$ and $j_{ab}(u)$ are density of energy and
angular momentum, respectively. One can also relate ${\cal A}_0$
to the density of the electric charge distribution $\rho$
\be
{\cal A}_0(u)= \kappa g_n \rho(u)\, .
\ee

Let us denote $b_{ab}=a_{ac}a_{bc}$ and
$b=\delta^{ab}b_{ab}=a_{ac}a_{ac}$. Then one has
\be\non
j_f=-{1\over 2}f_{ab}f^{ab}= {2b\over r^{2n}}+{n(n-4) b_{ab}x_a x_b\over
r^{2n+2}}\, .
\ee
\be\non
j_{{\cal A}}=(\nabla {\cal A})^2={(n-2)^2{\cal A}_0^2\over r^{2n-2}}\, .
\ee
Instead of using \nn{psi} and \nn{gfn}, one can determine $\psi$ by
directly solving the Laplace equation.
We write $\psi=\psi_f+\psi_{{\cal A}}$ as a sum  of two
solutions of \nn{J} with the sources $j_{f}$ and $j_{{\cal A}}$,
respectively. By using the  relations
\be\non
\lap \left( {1\over r^{2m}}\right)= {2m(m+1)\over r^{2m+2}}\, ,
\ee
\be\non
\lap \left( {x_a x_b \over r^{2m}}\right)=2m(m-2){x_a x_b \over
r^{2m+2}}+2{\delta_{ab}\over r^{2m}}\, ,
\ee
one can check that
\be
\psi_f={\alpha \over r^{2n-2}}+{\beta_{ab}x_a x_b\over r^{2n}} \hh
\psi_{{\cal A}}={(n-2)\over 2(n-1)} {{\cal A}_0^2\over r^{2n-4}}\, .
\ee
Here
\be
\alpha={b\over 2(n-1)(n-2)}\hh
\beta_{ab}={n-4\over 2(n-2)} b_{ab}\, ,
\ee

\section{Four-Dimensional Charged Gyratons}

The four-dimensional case is degenerate and requires a special
consideration. In the lowest order of the harmonic decomposition one
has
\be
\varphi=-{\kappa \over \pi\sqrt{2}} \varepsilon(u)\ln r  \hh
{\cal A}=-{\kappa \over 2\pi} \rho(u) \ln r \hh
a_a={\kappa j(u)\over 2\pi} {\epsilon_{ab} x^b\over r^2}\, . 
\ee
Simple calculations give
\be\non
j_f=0\hh
j_{ {\cal A}}={\kappa^2\over 4\pi^2}\rho^2(u){1\over r^2}.
\ee
To obtain $\psi_{{\cal A}}$ we use the relation ($p=p(r)$)
\be\non
\lap p={d^2p\over dr^2}+{n-1\over r}{dp\over dr}\, .
\ee
Solving equation $\lap \psi_{ {\cal A}}=j_{ {\cal A}}$ one finds
\be
\psi_{ {\cal A}}={\kappa^2\over 8\pi^2}\rho^2(u) \ln^2(r)\, .
\ee
To summarize, the metric of the $4-$dimensional charged gyraton can be
written in the form
\be \n{4d}
ds^2=-2~du~dv+dr^2+r^2 d\phi^2 +{\kappa\over 2\pi} j(u) du d\phi 
+ \Phi du^2\, ,
\ee
\be
\Phi= -{\kappa \sqrt{2}\over 2\pi} \varepsilon(u) \ln r +{\kappa^2\over 8\pi^2}
\rho^2(u)\ln^2(r)
\, .
\ee
This solution can be easily generalized to the case when the energy
density distribution has higher harmonics. It is sufficient to add to
$\Phi$ the function 
\be\n{ad}
\varphi'=\sum_{n\ne 0} {b_n \over r^{|n|}}
e^{in\phi}\hh
\bar{b}_{n}(u)=b_{-n}(u)\, .
\ee

The 4D solution of Eintein-Maxwell equations corresponding to a cylindrical
non-rotating charged source moving with the speed of light was
found in 1970 by Bonnor \cite{Bo2}. He also obtained \cite{Bo3} the 4D solution for the
gravitational field a neutral spinning null fluid. These solutions outside the
source correspond to our formulas for a particular choice of
charge and angular momentum distributions.


\section{Summary and discussions}

In the present paper we obtained solutions of Einstein-Maxwell equations for
the relativistic charged gyratons. These solutions generalize the results of
\cite{FrFu:05,FrIsZe:05} to the case when in addition to the finite energy and
angular momentum the gyraton has an electric charge. The parameters specifying
the solutions are arbitrary functions of $u$.  The obtained solutions have an
important property that all scalar invariants constructed from the Riemann
curvature, electric field strength and their covariant derivatives vanish. In
the case of shock waves this property was the reason why they do not receive
quantum  and $\alpha'$ corrections \cite{HorowitzSteif:90,AmatiKlimcik:89}. So,
one can expect that the gyraton metric also does not have quantum corrections
and, hence, it is probably the exact solution of the quantum problem too.  We
demonstrated that for the charged gyratons the Einstein-Maxwell field equations
in $D$-dimensional spacetime reduce to a set of linear equations in the
Euclidean $(D-2)$-dimensional space. This property of gyraton solutions is
similar to the properties of charged M-branes in string theory \cite{Maeda:05}
\footnote{We are grateful to  Kei-ichi Maeda for this remark.}.  In the absence of
the charge the gyraton solutions can be generalized to the case of
asymptotically AdS spacetimes \cite{FrZelADS:05}. It would be interesting to
find a similar generalization for charged gyratons. In the supergravity theory
there are other charged objects similar to gyratons, where, for example,  the
Kalb-Ramond field appears instead of the Maxwell field. It happens that in the
the Brinkmann metric ansatz the supergravity equations can also be reduced to a
system of linear equations and solved exactly \cite{FrLin:05}.

\noindent 
\section*{Acknowledgments} 
\noindent The authors are
grateful to Robert Mann for stimulating discussion. This work was
supported by the Natural Sciences and Engineering Research Council of
Canada and by the Killam Trust.


\end{document}